\documentclass[
reprint,
twocolumn,
bibnotes,
amsmath,amssymb,
]{revtex4-2}
\usepackage{subcaption}
\usepackage{tikz}
\usepackage[font=footnotesize]{caption}
\usepackage{mwe}
\usepackage{graphicx}
\usepackage{dcolumn}
\usepackage{bm}
\usepackage{hyperref}

\DeclareRobustCommand{\orcidicon}{%
	\begin{tikzpicture}
		\draw[lime, fill=lime] (0,0)
		circle [radius=0.16]
		node[white] {{\fontfamily{qag}\selectfont \tiny ID}};
		\draw[white, fill=white] (-0.0625,0.095)
		circle [radius=0.007];
	\end{tikzpicture}
	\hspace{-2mm}
}

\foreach \x in {A, ..., Z}{%
	\expandafter\xdef\csname orcid\x\endcsname{\noexpand\href{https://orcid.org/\csname orcidauthor\x\endcsname}{\noexpand\orcidicon}}
}

\DeclareRobustCommand{\orcidicon}{%
	\begin{tikzpicture}
		\draw[lime, fill=lime] (0,0)
		circle [radius=0.16]
		node[white] {{\fontfamily{qag}\selectfont \tiny ID}};
		\draw[white, fill=white] (-0.0625,0.095)
		circle [radius=0.007];
	\end{tikzpicture}
	\hspace{-2mm}
}

\foreach \x in {A, ..., Z}{%
	\expandafter\xdef\csname orcid\x\endcsname{\noexpand\href{https://orcid.org/\csname orcidauthor\x\endcsname}{\noexpand\orcidicon}}
}

\begin{document}
	
	
	\title{Topological phase transition in monolayer 1T$^\prime$-MoS$_2$}

	\author{Mohammad Mortezaei Nobahari\orcidA}
	
    \author{Mahmood Rezaei Roknabadi\orcidB}

	\affiliation{
		Department of Physics, Ferdowsi University of Mashhad, Iran}
	\author{\href{mortezaie.mm71@gmail.com}{mortezaie.mm71@gmail.com} 
		\\ \href{roknabad@um.ac.ir}{roknabad@um.ac.ir} }
	
	
	
	
	\date{\today}
	\begin{abstract}
	1T$^{\prime}$ phase of the monolayer transition metal dichalcogenides has recently attracted attention for its potential in nanoelectronic applications. We theoretically prove the topological behavior and phase transition of 1T$^{\prime}$-MoS$_2$ using $k.p$ Hamiltonian and linear response theory. The spin texture in momentum space reveals a strong spin-momentum locking with different orientations for the valence and conduction bands. Also, Berry curvature distributions around the Dirac points highlight the influence of $\alpha$ parameter demonstrating a topological phase transition in 1T$^\prime$-MoS$_2$. For $\alpha<1$ the spin Hall conductivity is the only non-zero term $(C_s=1$   and $C_v=0)$, corresponding to a quantum spin Hall insulator (QSHI) phase, while for $\alpha>1$, valley Hall conductivity prevails, indicating a transition to a band insulator (BI). Further analysis explores the spin-valley-resolved Hall conductivity and Chern numbers across varying values of $\alpha$, $V$, and Fermi energy, uncovering regions of  non-trivial  and trivial topological phases (TTP) and the role of the edge modes. The zero total Nernst coefficient across energy ranges suggests strong cancellation between spin and valley contributions, providing insights into the material's potential for thermoelectric applications and spintronic devices.
		
	\end{abstract}
	
	\maketitle
	
	
	\section{\label{sec:1}Introduction }

In condensed matter physics, the rise of topological materials has marked a significant shift in research, unveiling quantum phenomena with far-reaching consequences. Among these materials, quantum spin Hall insulators (QSHI) play a crucial role, representing a transformative advancement in our understanding of topologically nontrivial electronic states~\cite{RevModPhys.82.3045}.

The concept of a QSHI was initially introduced by Bernevig et al.~\cite{PhysRevLett.96.106802}, signifying a groundbreaking departure from traditional electronic behavior by incorporating topological protection for electronic states. These materials exhibit insulating characteristics in their bulk while supporting stable conducting edge states, paving the way for dissipationless electronic transport and innovative spin-based applications~\cite{RevModPhys.83.1057,Hsieh2009,Bercioux_2015}.

Experimental studies have confirmed the presence of QSHI behavior across various material platforms, including one- and two-dimensional systems as well as engineered heterostructures, broadening the scope of potential applications for these topological electronic states~\cite{doi:10.1126/science.1148047,Yan_2012,Brune2012}. These investigations have shed light on the complex relationship between topological and electronic properties at the core of QSHIs.

Monolayer transition metal dichalcogenides (TMDCs) in the T$^\prime$ structural phase, characterized by the general chemical formula MX$_2$ with M representing W or Mo and X representing Te, Se, or S, have been theoretically predicted to exhibit properties of a QSHI~\cite{doi:10.1126/science.1256815} and attracted attentions~\cite{PhysRevLett.108.196802,Wang2012,Xu2014,Manzeli2017,doi:10.1126/science.aan6003,PhysRevLett.124.166803,Tang2017,Das2020}. 1T$^\prime$-MoS$_2$ has emerged as a prominent two-dimensional (2D) material, distinguished by its unique structural and electronic properties. Unlike its more commonly studied counterpart, 2H-MoS$_2$, which exhibits a stable semiconducting phase, 1T$^\prime$-MoS$_2$ adopts distorted octahedral coordination that results in a metallic character and significant potential for various applications in electronics, catalysis, and energy storage. The transition from the 2H to the 1T$^\prime$ phase can be induced through methods such as chemical intercalation or strain engineering, highlighting the material's versatility and tunability~\cite{C6TA09409K,PhysRevB.109.144421}.
	
The unique electronic structure of 1T$^\prime$-MoS$_2$ is characterized by the presence of helical edge states and strong spin-orbit coupling, which can lead to interesting phenomena such as spin polarization and topological properties~\cite{Luo2020}. These features make 1T$^\prime$-MoS$_2$ a candidate for spintronic applications and quantum computing, where control over electron spin and charge is essential~\cite{D0NJ03788E}. Furthermore, its high surface area and catalytic activity have positioned it as a promising material for electrocatalytic applications, including hydrogen evolution reactions.

The significance of QSHIs extends beyond theoretical physics, reaching into practical applications in electronics and spintronics. The chiral nature of the edge states in QSHIs presents an exciting opportunity for dissipationless spin transport, which could lead to the development of efficient spin-based logic and memory devices that utilize the spin degrees of freedom of electrons~\cite{Pesin2012,Jansen2012,Manchon2015}. Additionally, the complex interplay between the topological and electronic properties of these materials underlies their potential for enabling topologically protected quantum computation and information processing~\cite{KITAEV20032,RevModPhys.80.1083}.

In light of these developments, the experimental realization of the quantum spin Hall effect and the identification of materials with topologically nontrivial electronic states have created unique opportunities to exploit their extraordinary properties~\cite{Xu2012,PhysRevLett.113.137201}. The synergy between theory and experiment has propelled the field of topological electronics into an era of unprecedented promise and potential.

Although the topological behavior of 1T$^{\prime}$-MoS$_2$ has been predicted previously, the Hall conductivity and Chern numbers of this material have not been investigated theoretically so far. In this study, we prove this topological behavior using $k.p$ Hamiltonian and Kubo formula.\par

This paper begins by exploring the theoretical background in section \ref{sec:2} to gain insight into the properties of 1T$^\prime$-MoS$_2$. Next, theoretical frameworks are applied to calculate these properties in section \ref{sec:3}, and the results are summarized in section \ref{sec:4}.

\begin{figure}[]                                                                                                                                                          
\includegraphics[width=0.5\textwidth]{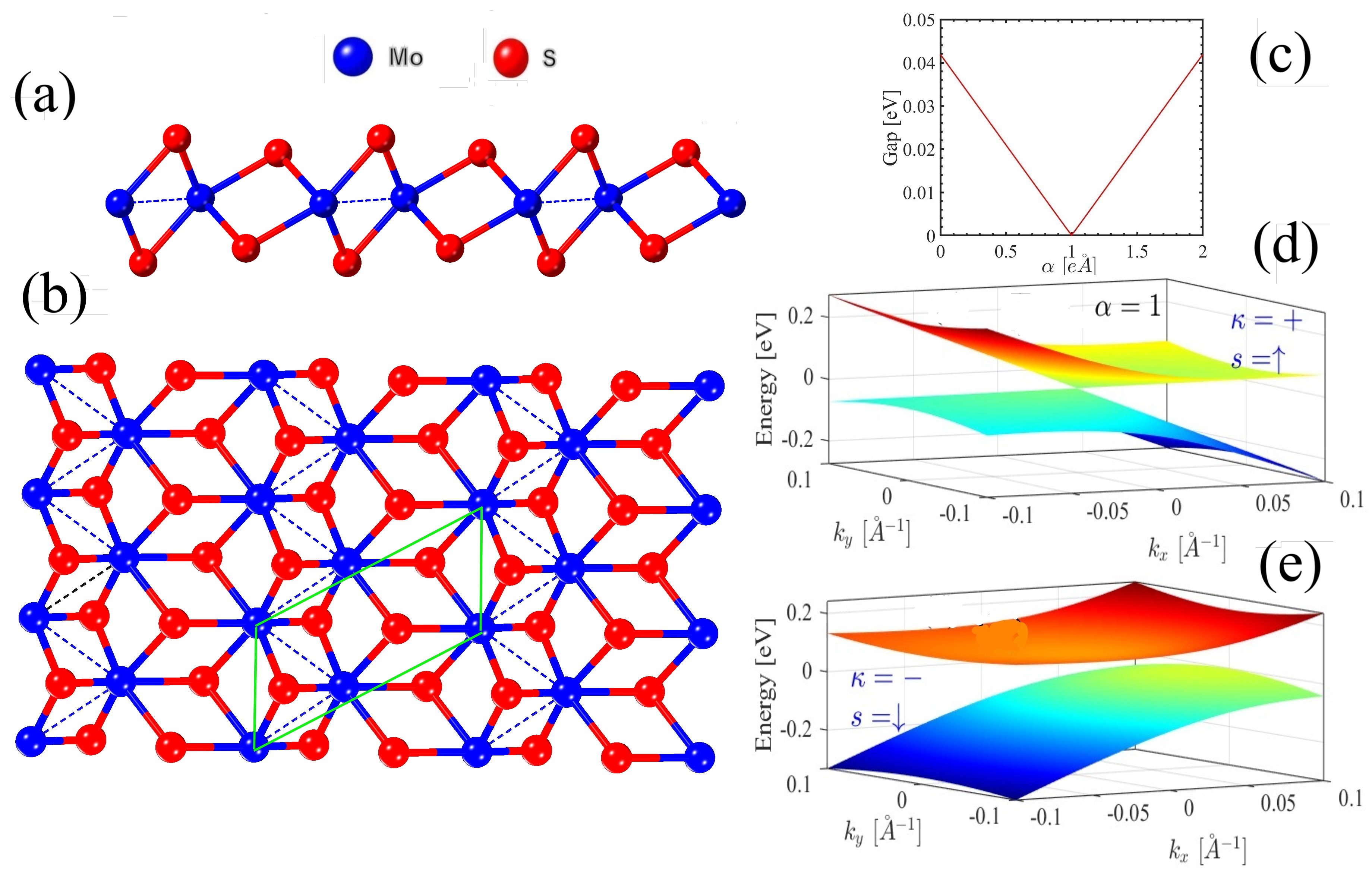}
\caption{(a) The top view of monolayer 1T$^{\prime}$-MoS$_2$, consisting of Mo and S atoms represented in blue and red colors respectively. The unit cell consists of four atoms and is represented by the green rectangle. Panel (b) displays the relationship between the spin-orbit coupling gap and $\alpha$. Panels (c) and (d) illustrate the 3D band structure for $\alpha=1$, with $\kappa=+$ and spin $s=\uparrow$, as well as for $\kappa=-$ and spin $s=\downarrow$ respectively.}\label{fig:1}                                                            
\end{figure}

     
     
 
     
    



\section{Theory\label{sec:2}}
Figures \ref{fig:1} (a) and (b) present the top and side views of 1T$^{\prime}$-MoS$_2$, which is composed of two types of atoms: molybdenum (Mo) represented in blue and sulfur (S) shown in red. The unit cell, outlined by a green rectangle, contains four atoms. The low-energy $k.p$ Hamiltonian of monolayer 1T$^{\prime}$-MoS$_2$ in the $x-y$ plane and in presence of a perpendicular electric field $E_z$ and gate voltage V is well addressed before~\cite{doi:10.1126/science.1256815,Das2020,PhysRevB.107.035301}. The Hamiltonian can be written as a summation of the $H_{k.p}$ and external perturbations.

\begin{equation}
 H=H_{k.p}+H_{E_z}+V
 \label{eq:1}
\end{equation}

where 

\begin{equation}
H_{k.p} = \left({\begin{array}{cccc} E_p& 0& -iv_1\hbar q_y &v_2\hbar q_y\\
0 & E_p &v_2\hbar q_y & -iv_1\hbar q_x\\ iv_1\hbar q_x & v_2\hbar q_y&E_d & 0\\
v_2\hbar q_y& iv_1\hbar q_x & 0 & E_d\\\end{array}}\right)\label{eq:2}
\end{equation}

and 

\begin{equation}
H_{E_z} = \alpha\Delta_{so}\left({\begin{array}{cccc} 0& 0& 1 &0\\
0 & 0 &0 & 1\\ 1  & 0&0 & 0\\
0& 1 & 0 & 0\\\end{array}}\right)\label{eq:3}
\end{equation}

Here, V is a $4 \times 4$ diagonal matrix, and $\Delta_{so} = 41.9$meV  represents the fundamental spin-orbit coupling gap at the Dirac points $(0, \kappa\Lambda)$. The parameter $\alpha$ is defined as $|E_z/E_c|$, where $E_z$ is the perpendicular electric field and $E_c  = 1.42$V/nm$^{-1}$ is the critical electric field required for a topological phase transition. The on-site energies for the $p$ and $d$ orbitals are given by the $E_p=\delta_p + \frac{\hbar^{2}q_x^2}{2m_x^p} + \frac{\hbar^2q_y^2}{2m_y^p}$ and $E_d = \delta_d + \frac{\hbar^2 q_x^2}{2m_x^d} + \frac{\hbar^2 q_y^2}{2m_y^d}$ respectively, where $q_{x,y}$ is the electron momentum. The parameters are defined as follows:  $\delta_p = 0.46$eV, $\delta_d = -0.20$eV, $m_x^p = -0.50m_0$, $m_y^p = -0.16m_0$, $m_x^d = 2.48m_0$, $m_y^d = 0.37m_0$, and $m_0$ is the free electron mass, $\nu_1 = 3.87\times 10^5m/s$ and $\nu_2 = 0.46 \times 10^5m/s$ are the Fermi velocities along the $x$ and $y$ directions respectively.

Using a unitary transformation the Hamiltonian reads as

\begin{equation} 
	\begin{split}
		H_{\kappa,s}(k)&=\hbar k_x\nu_1\sigma_y-\hbar k_y(s\nu_2\sigma_x+\kappa\nu_{-}\sigma_0 \\ 
		&+\kappa\nu_{+}\sigma_z)+\Delta_{so}(\alpha-s\kappa)\sigma_x+V\sigma_{0}\label{eq:4}
	\end{split}
\end{equation}

Where $s=\pm$ is related to the spin up and down respectively, and $\kappa=\pm$ represents valley indexes ($K$ and $K'$). The tilting velocities are $\nu_{-}= \frac{\hbar q_0}{2}(-\frac{1}{m_y^p}-\frac{1}{m_y^d}) = 2.84 \times 10^5m/s$ and $\nu_{+}= \frac{\hbar q_0}{2}(-\frac{1}{m_y^p}+\frac{1}{m_y^d}) = 7.18 \times 10^5m/s$. 

After diagonalization of the Hamiltonian, the eigen energies are

\begin{align}
	\varepsilon_{\kappa,s}^{\pm}(k_x,k_y)&=-\hbar k_y\kappa\nu_{-}+V \nonumber\\
	&\pm\sqrt{[\zeta_{\kappa,s}(k_y)]^2+[\Lambda(k_y)]^2-[\eta(k_x)]^2}\label{eq:5}
\end{align}

and corresponding eigenstates are
\begin{equation}
\psi_{\kappa,s}^{\pm}(k_x,k_y)=\begin{pmatrix}
\frac{\mp \eta(k_x)\mp \zeta_{\kappa,s}(k_y)}{\pm \Lambda_{\kappa}(k_y)+\sqrt{[\zeta_{\kappa,s}(k_y)]^2+[\Lambda(k_y)]^2-[\eta(k_x)]^2}}\\
1 \label{eq:6}
\end{pmatrix}
\end{equation}

where 
\begin{equation}
 \zeta_{\kappa,s}(k_y)=\hbar k_ys\nu_2-\Delta_{so}(\alpha-s\kappa)\label{eq:7}
\end{equation}

\begin{equation}
 \Lambda_{\kappa}(k_y)=\hbar k_y\kappa\nu_{+}\label{eq:8}
\end{equation}

\begin{equation}
 \eta(k_x)=i\hbar k_x\nu_1\label{eq:9}
\end{equation}

The 3D band structure of 1T$^{\prime}$-MoS$2$ is depicted in Figs.~\ref{fig:1} (c) and (d) under the influence of an electric field ($\alpha \neq 0$). When $\alpha = 1$, $\kappa = +$, and $s=\uparrow$, the spin-orbit gap is closed (see Fig.~\ref{fig:1} (b)). Conversely, for $\kappa = -$ and $s = \downarrow$, a small gap of approximately 0.04eV is observed. As we will demonstrate later, this critical amount of the $\alpha$ is a key factor in the topological phase transition.

Pseudo-spin texture in 1T$^{\prime}$-MoS$_2$ emerges from its unique electronic structure, particularly due to the presence of two distinct valleys in the Brillouin zone, often referred to as the $K$ and $K^{\prime}$ valleys. These valleys can be treated as pseudo-spin states, where electrons in the $K$ valley are analogous to spin-up and those in the $K^{\prime}$ valley to spin-down. The two components of the pseudo-spin textures can be calculated by $S_x^{\pm}=\langle\psi_{\kappa,s}^{\pm}|\sigma_x|\psi_{\kappa,s}^{\pm}\rangle$ and $S_y^{\pm}=\langle\psi_{\kappa,s}^{\pm}|\sigma_y|\psi_{\kappa,s}^{\pm}\rangle$, where $\sigma_x$ and $\sigma_y$ are the Pauli matrices and $\pm$, stems for the pseudo-spin texture in the conduction and valence band respectively. After some algebra we have

\begin{equation}
 S_x^{\pm}=\mp\frac{2\zeta_{\kappa,s}(k_y)}{\pm\Lambda_{\kappa}(k_y)+\sqrt{[\zeta_{\kappa,s}(k_y)]^2+[\Lambda(k_y)]^2-[\eta(k_x)]^2}}\label{eq:10}
\end{equation}

and

\begin{equation}
 S_y^{\pm}=\pm\frac{2\hbar k_x\nu_1}{\pm\Lambda_{\kappa}(k_y)+\sqrt{[\zeta_{\kappa,s}(k_y)]^2+[\Lambda(k_y)]^2-[\eta(k_x)]^2}}\label{eq:11}
\end{equation}

The azimuthal angle of the pseudo-spin can be expressed as
\begin{equation}
\tan(\theta_s)=\frac{S_y}{S_x}=-\frac{\hbar k_x\nu_1}{\hbar k_ys\nu_2-\Delta_{so}(\alpha-s\kappa)}\label{eq:12}
\end{equation}

We calculate optical Hall conductivity using the Kubo formula and Berry curvatures. The Kubo formula for Hall conductivity is given by ~\cite{MortezaeiNobahari2023,RevModPhys.82.1959,Hajati_2023}                                                  
				
				\begin{equation}	                                                                                                                                                        
					\sigma_{\kappa,s}=\frac{e}{h}\int{f^n_{\kappa,s}(\vec{k})\Omega_{\kappa,s}^n(\vec{k})d^2k} \label{eq:13}	                                                                              
				\end{equation} 
				
				Where the Berry curvature and direction-dependent velocities operators are defined as

				\begin{equation}                                                                                                                                                          
					\Omega_{\kappa,s}^n(\vec{k})=-2\text{Im}\sum _{n\neq n^{\prime}} \frac{\beta_{\kappa,s}^{n,n^{\prime},x}(\vec{k})\beta_{\kappa,s}^{n^{\prime},n,y}(\vec{k})}{(\varepsilon_{\vec{k},\kappa,s}^{n}-\varepsilon_{\vec{k},\kappa,s}^{n^{\prime}})^2}.\label{eq:14}                                                                                      
				\end{equation}                                                        	
				\begin{equation}                                                                                                                                                         
					\beta_{\kappa,s}^{n,n^{\prime},x}(\vec{k})=\langle \vec{k}; n,\kappa,s |j_{x}| \vec{k}; n^{\prime},\kappa,s\rangle   \label{eq:15}                                                                             
				\end{equation}                                                                                                                                              
				
				\begin{equation}                                                                                                                                                          
					\beta_{\kappa,s}^{n^{\prime}, n, y}(\vec{k})=\langle \vec{k};n^{\prime},\kappa,s |j_y| \vec{k}; n, \kappa,s\rangle                                                                             
				\end{equation}\label{eq:16}                                                                                                                                               
				
The expression $f_{\kappa,s}^n(\vec{k})=[1+\text{exp}(\varepsilon_{\kappa,s}^n-E_F)/k_BT]^{-1}$ represents the Fermi-Dirac distribution function for the nth band, where $E_F$ denotes the Fermi energy. The index $n =\pm$ indicates the conduction and valence bands, respectively. Also, the quantities $j_x=e\partial{H_{\kappa,s}(\vec{k})}/\partial k_x$ , and $j_y=e\partial H_{\kappa,s}(\vec{k})/\partial k_y$ describe the direction-dependent velocities along the $x$ and $y$ axes.
				
The spin-valley-resolved Hall conductivity is expressed as  $\sigma_{\kappa,s}=e^2/hC_{\kappa,s}$, where $C_{\kappa,s}$ represents the spin-valley-resolved Chern number

\begin{equation}
 C_{\kappa,s}=\frac{1}{2\pi}\int{f_{\kappa,s}\Omega_{\kappa,s}^n(\vec{k})d^2k}\label{eq:17} 
\end{equation}

The spin and valley Chern numbers are given by $C_s=(C_{\uparrow}-C_{\downarrow})/2$ and $C_v=(C_{+}- C_{-})/2$, respectively, in which $C_{\uparrow(\downarrow)}=\sum_{\kappa}{C_{\kappa,\uparrow(\downarrow)}}$	and $C_{+(-)}=\sum_{s}{C_{+(-),s}}$.

Having Berry curvatures we can calculate spin-valley Nernst coefficient as~\cite{PhysRevLett.115.246601,10.1063/1.4950854}

\begin{equation}
 \Gamma_{\kappa,s}^n=\frac{ek_B}{2\pi\hbar}\sum_n\int{\Omega_{\kappa,s}^n(\vec{k})S_{\kappa,s}^n(\vec{k})dk^2}\label{eq:18} 
\end{equation}

 where the entropy density is given by $S_{\kappa,s}^n(\vec{k})=-f_n(\vec{k})\ln{f_n(\vec{k})}-[1-f_n(\vec{k})]\ln[1-f_n(\vec{k})]$. The TNC can be calculated by summing over all spin-valley
resolved NCs $(\Gamma_{\uparrow}+\Gamma_{\downarrow}+\Gamma_{+}+\Gamma_{-})$ and the spin SNC and VNC are defined as $\Gamma_s=\Gamma_{\uparrow}-\Gamma_{\downarrow}$ and $\Gamma_v=\Gamma_+-\Gamma_-$ respectively.

\begin{figure}[h]
	
	\begin{tabular}{cc}
		\includegraphics[width=0.5\linewidth]{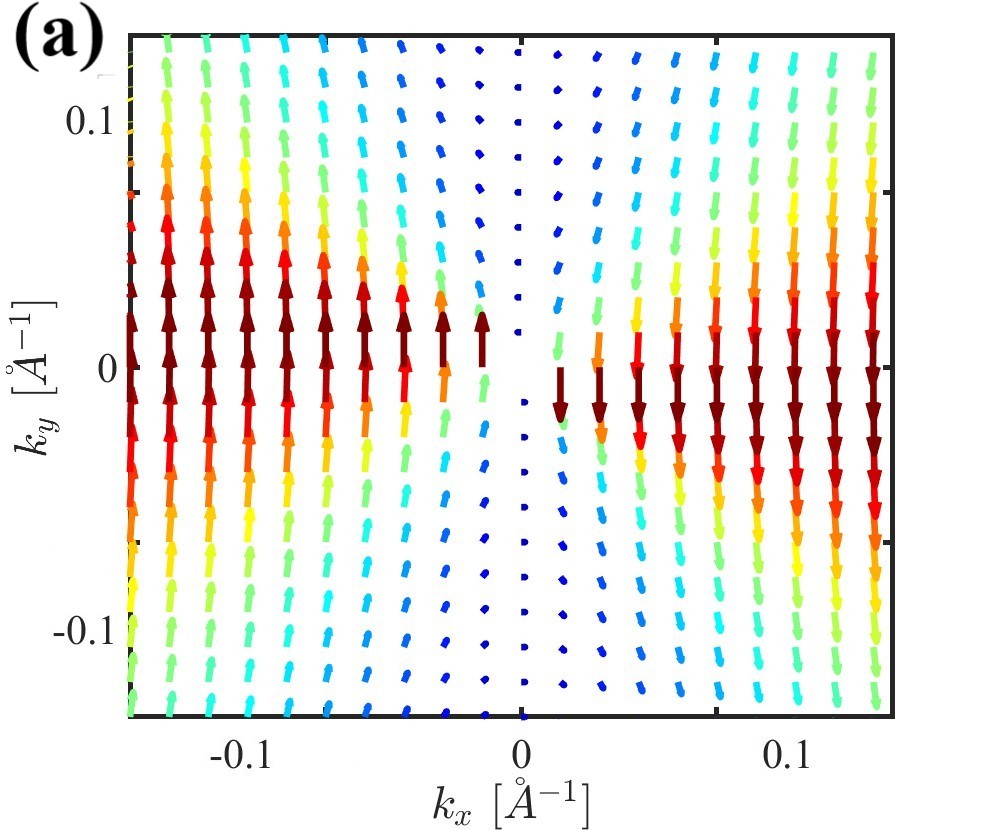}&
		\includegraphics[width=0.5\linewidth]{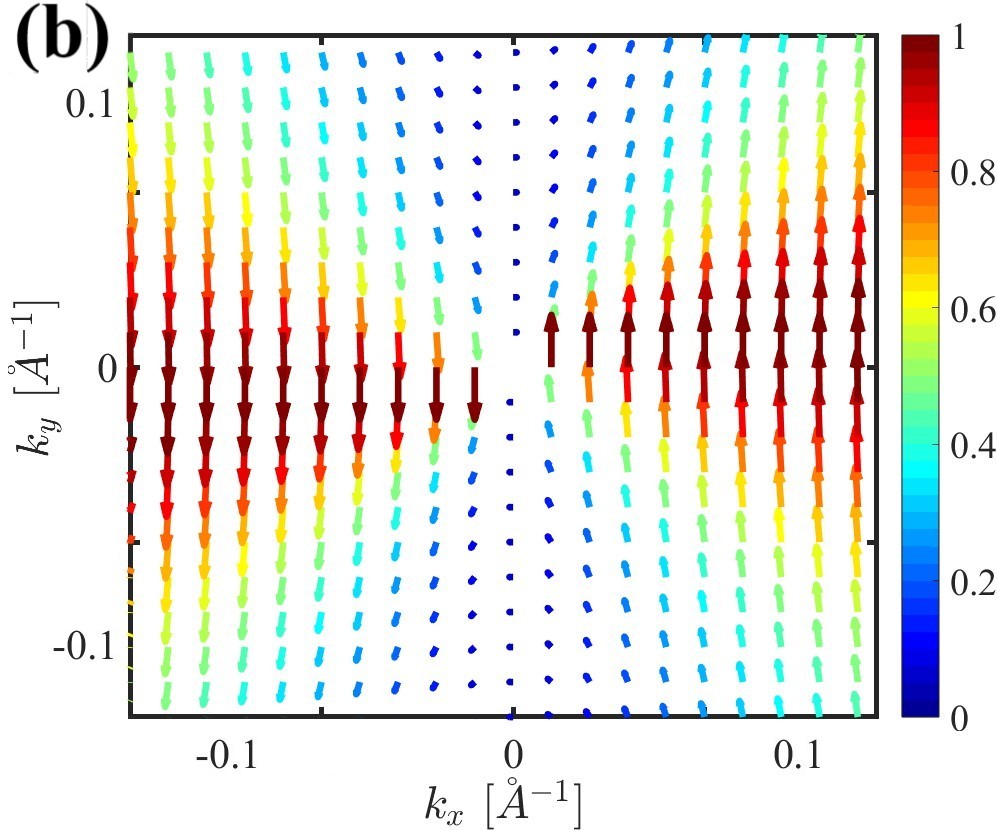}\\[2\tabcolsep]
	
	\end{tabular}
	\caption{Pseudo-spin texture in the momentum space for (a) conduction band and (b) valence band.}\label{fig:2}
\end{figure}

\begin{figure}[]
	\centering
	\begin{tabular}{cc}
		\includegraphics[width=0.5\linewidth]{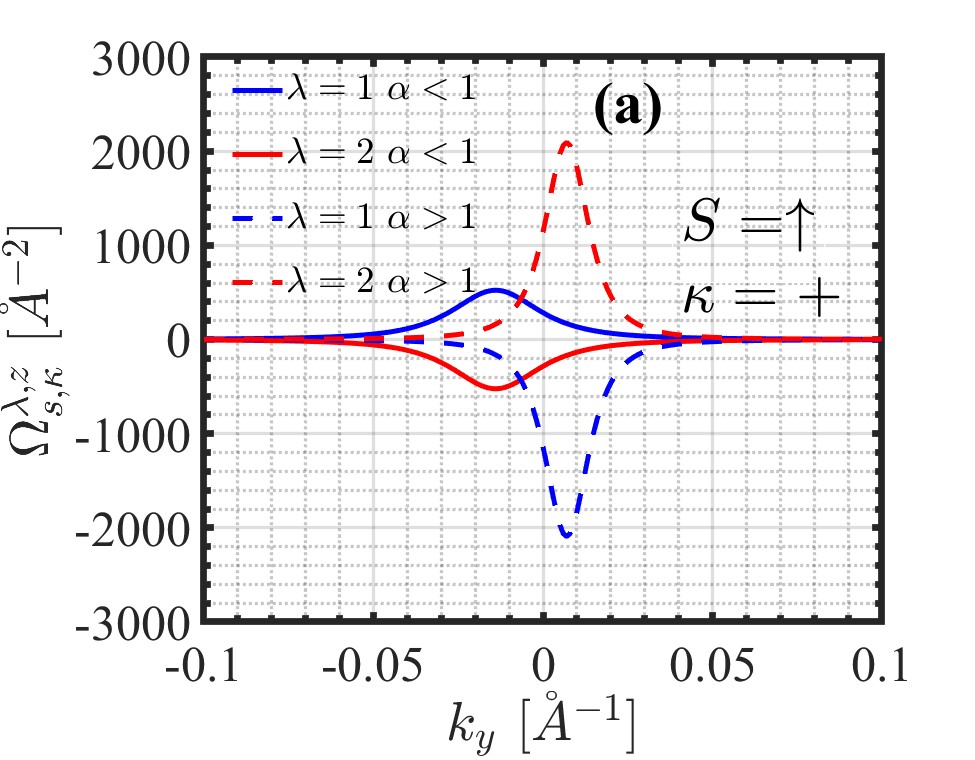}&
		\includegraphics[width=0.5\linewidth]{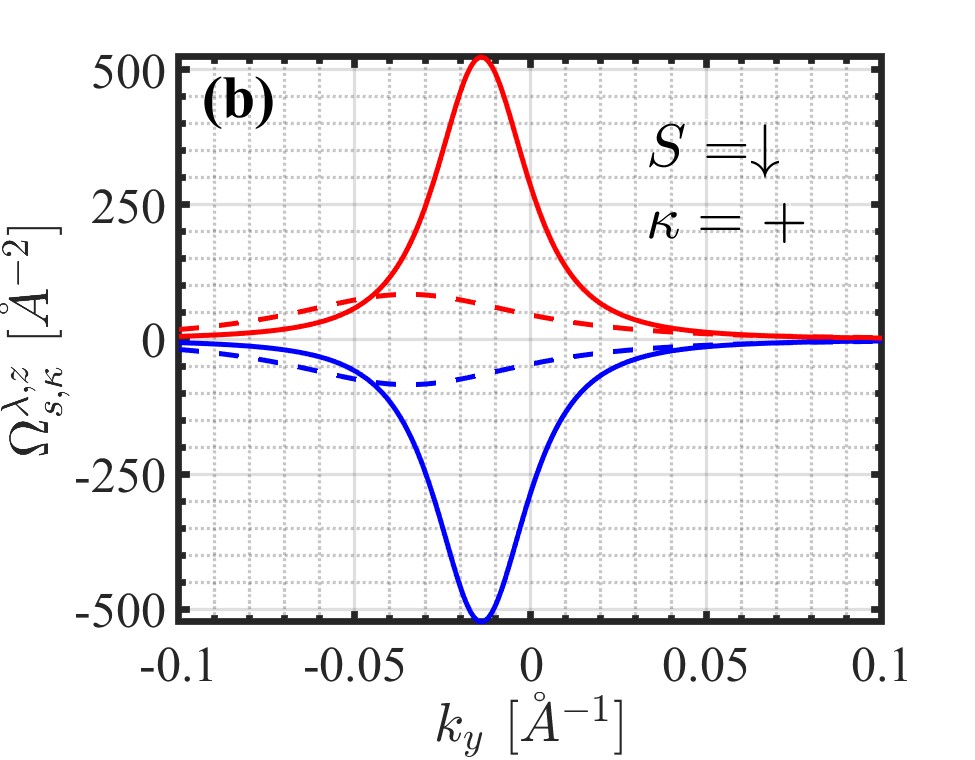}\\[2\tabcolsep]
		\includegraphics[width=0.5\linewidth]{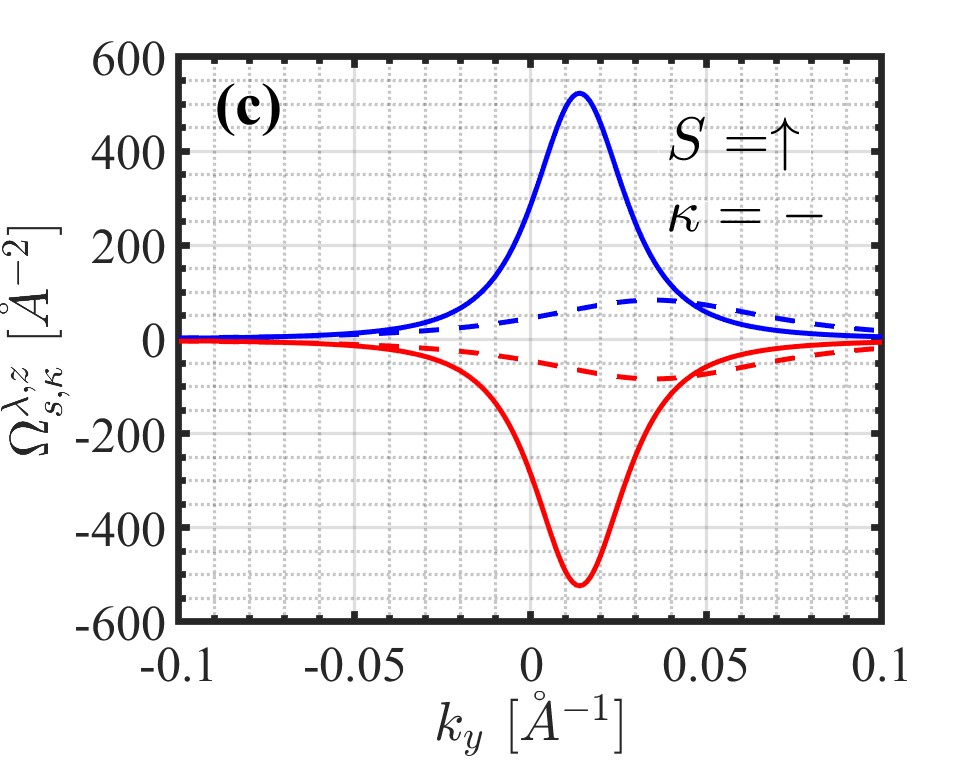}&
		\includegraphics[width=0.5\linewidth]{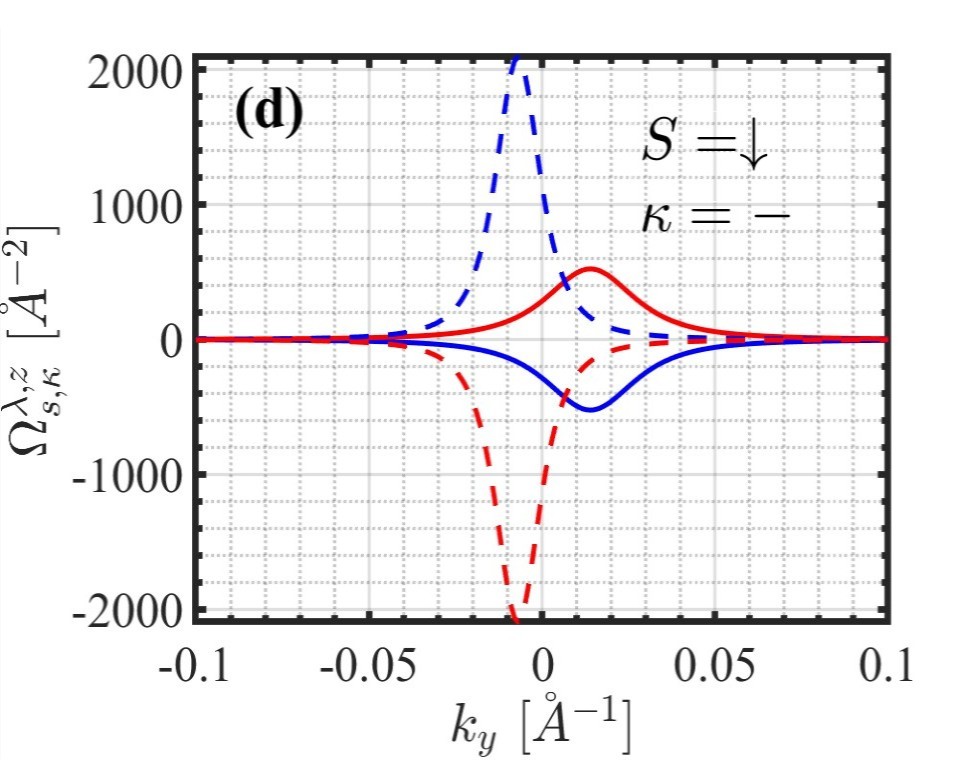}
	\end{tabular}
	\caption{Conduction and valence Berry curvature around the Dirac points for $\alpha=0$ and $\alpha=1.5$ in (a) $\kappa=+$, $s=\uparrow$, (b) $\kappa=+$, $s=\downarrow$, (c) $\kappa=-$, $s=\uparrow$, and (d) $\kappa=-$, $s=\downarrow$. $\lambda=1$ and $\lambda=2$ refer to the valence and conduction bands respectively.}\label{fig:3}
\end{figure}

\begin{figure}[]                                                                                                                                                          
\includegraphics[width=0.5\textwidth]{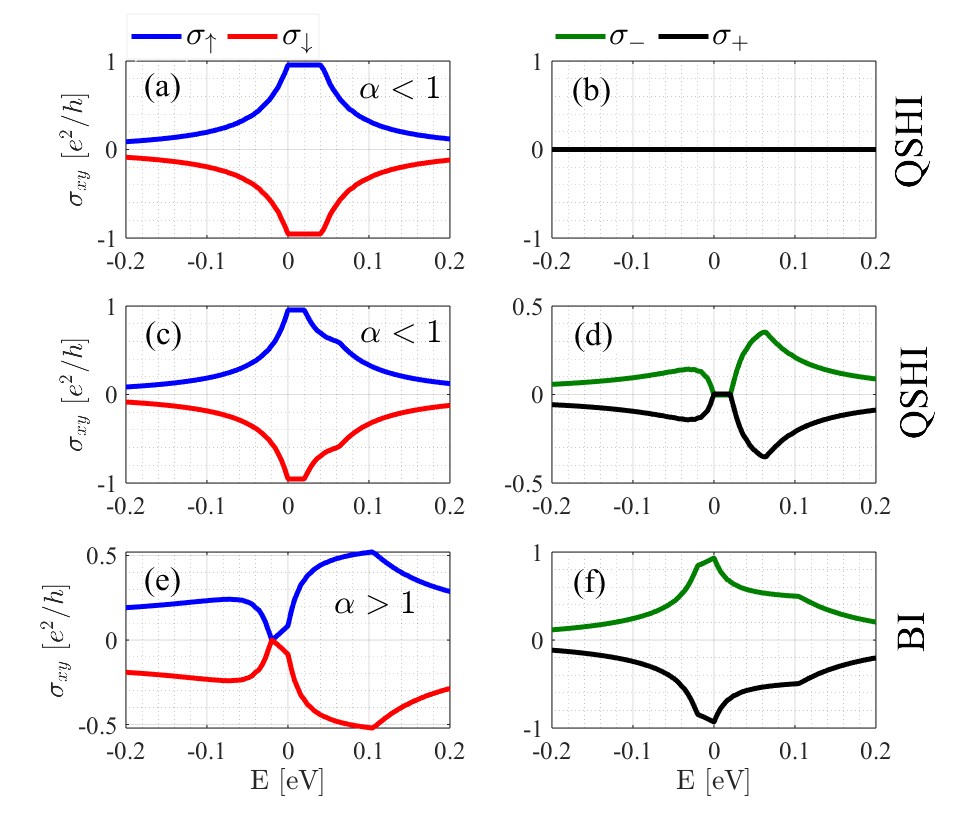}
\caption{Spin and valley Hall conductivities with respect to the Fermi energy at $T = 0$ K for $\alpha<1$ and $\alpha> 1$.}
	\label{fig:4}
	                                                          
\end{figure}

\begin{figure}[h]
	\begin{tabular}{cc}
		\includegraphics[width=0.5\linewidth]{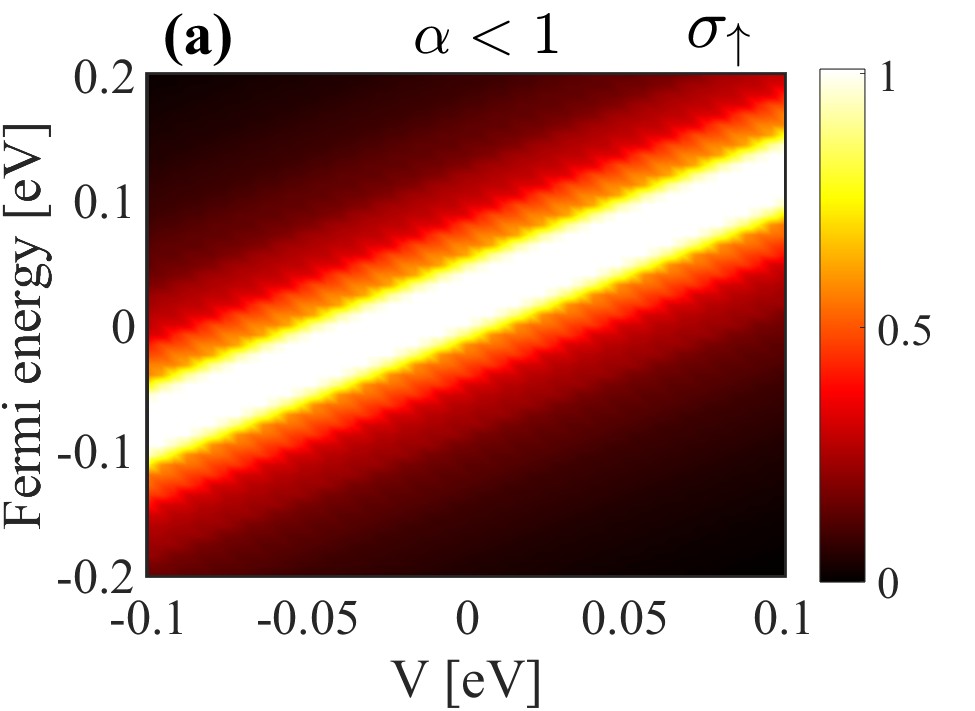}&
		\includegraphics[width=0.5\linewidth]{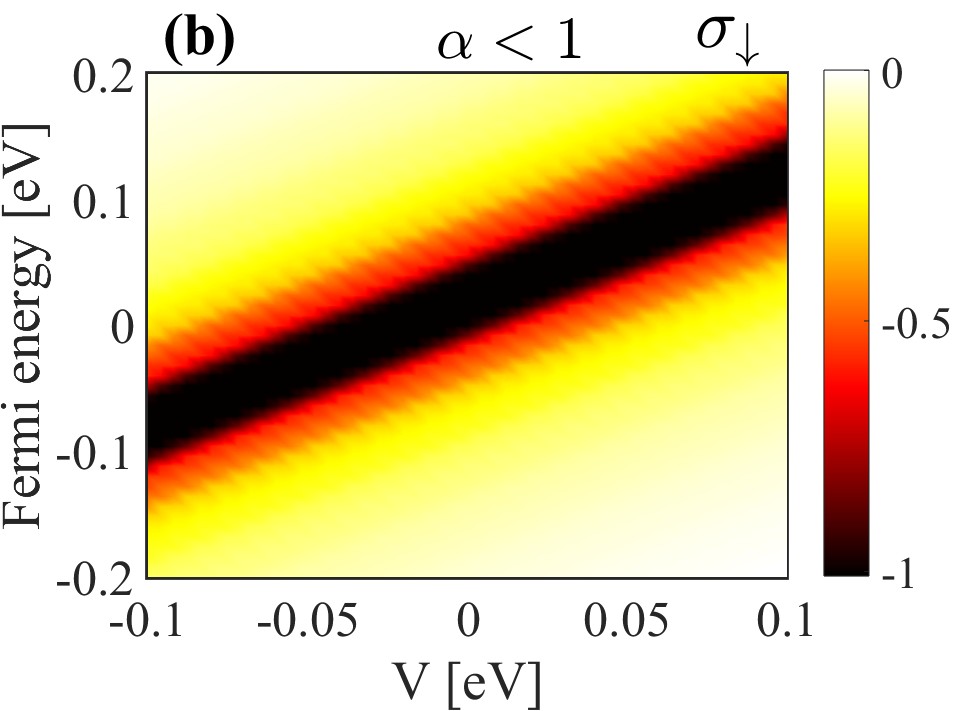}\\[2\tabcolsep]
		\includegraphics[width=0.5\linewidth]{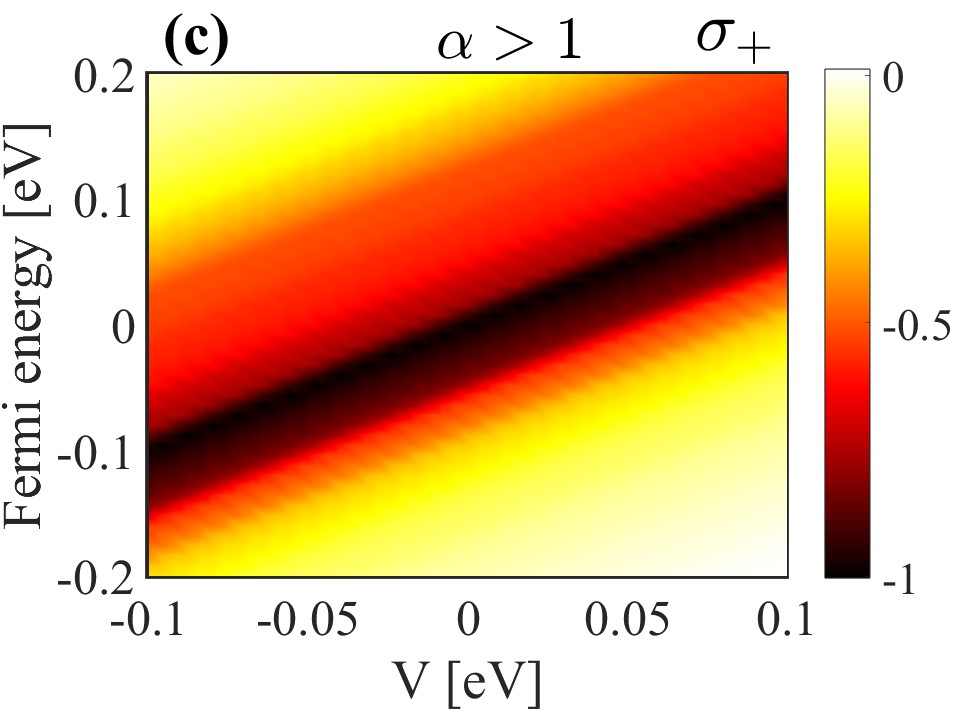}&
		\includegraphics[width=0.5\linewidth]{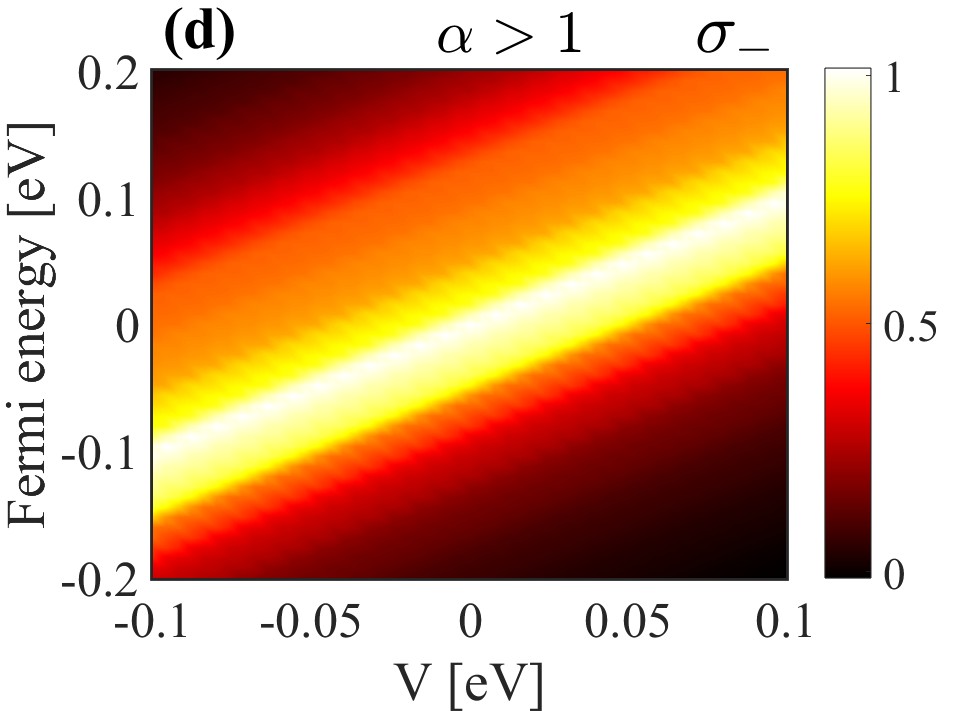}
	\end{tabular}
	\caption{The color density of the spin-valley-resolved Hall conductivity versus $V$ and the Fermi energy. Panel (a) and (b) show spin Hall conductivity for spin-up and spin-down respectively while $\alpha<1$, (c) and (d) is valley Hall conductivity for $\kappa=\pm$ respectively for $\alpha>1$.}\label{fig:5}
\end{figure}

\begin{figure}
	\begin{subfigure}{0.35\textwidth}
		\includegraphics[width=\textwidth]{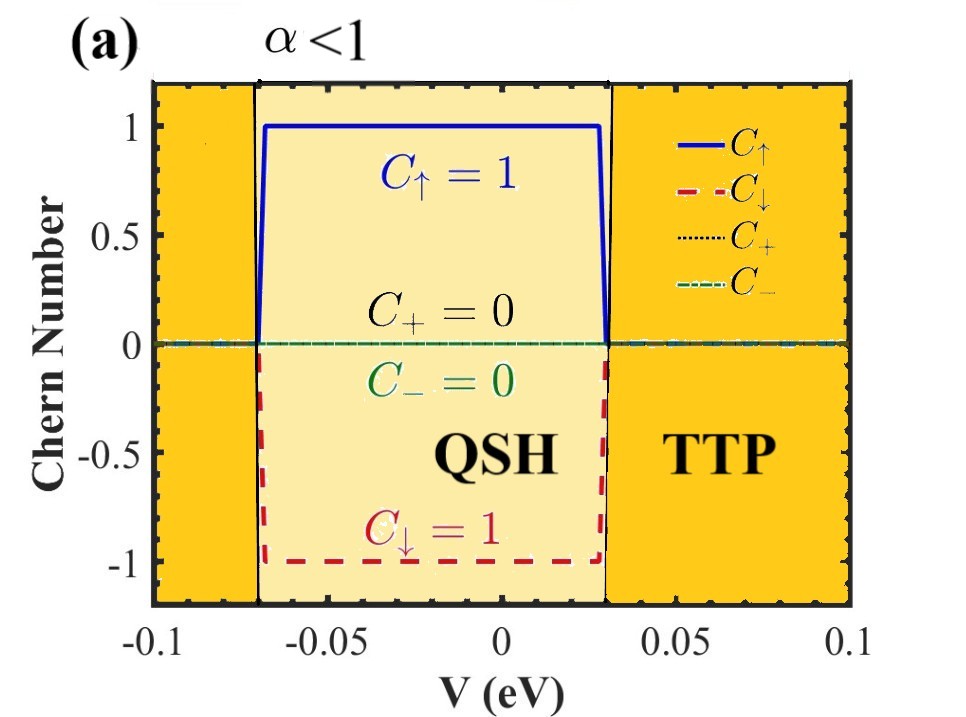}
		
		\label{fig:6a}
	\end{subfigure}
	\hfill
	\medskip
	\begin{subfigure}{0.35\textwidth}
		\includegraphics[width=\textwidth]{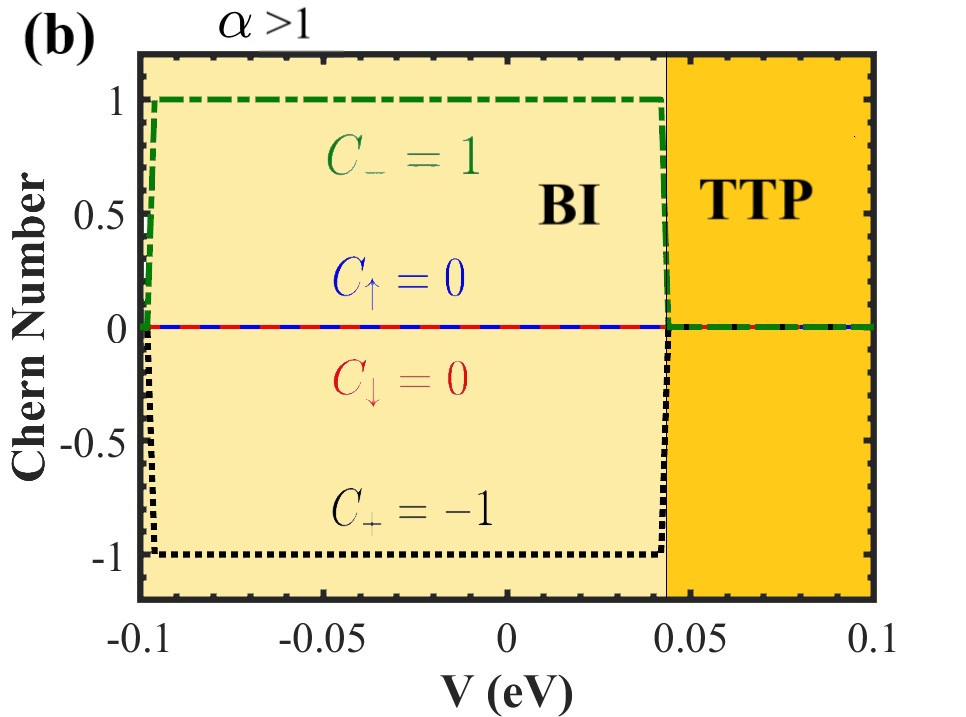}
		
		\label{fig:6b}
	\end{subfigure}
	\hfill
	\caption{The Chern number versus $V$ in the presence of the electric field for (a) $\alpha<1$ and (b) $\alpha>1$. The colored areas show the topological phase of the system.}
	\label{fig:6}
	\end{figure}

\begin{figure}[h]
	\begin{tabular}{cc}
		\includegraphics[width=0.5\linewidth]{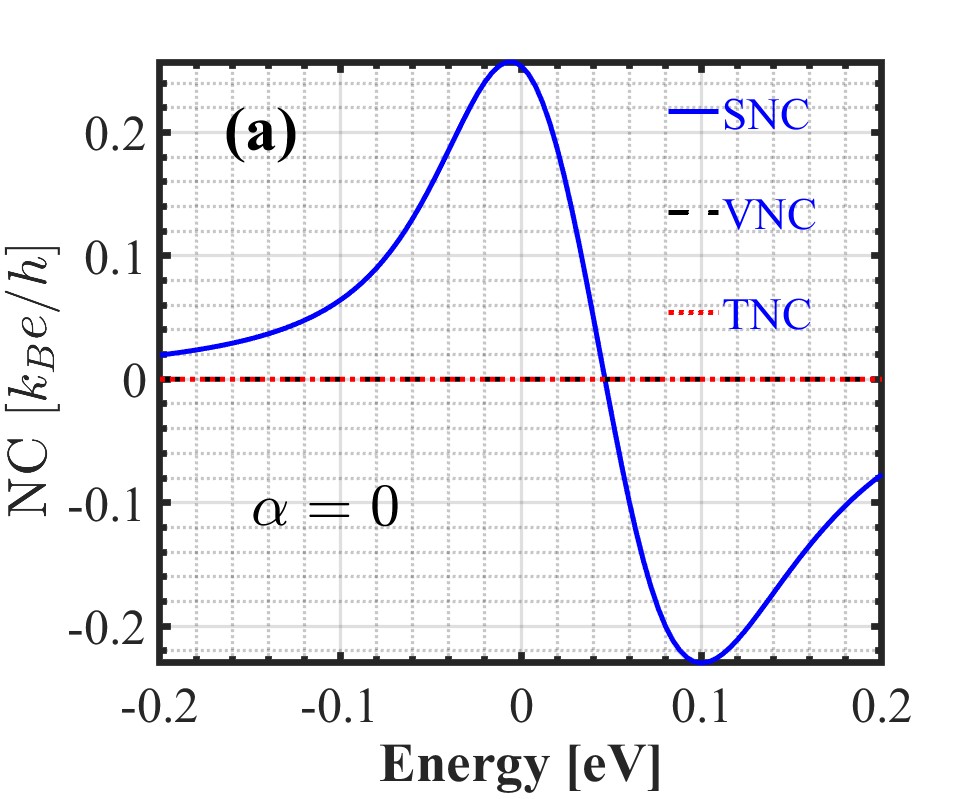}&
		\includegraphics[width=0.5\linewidth]{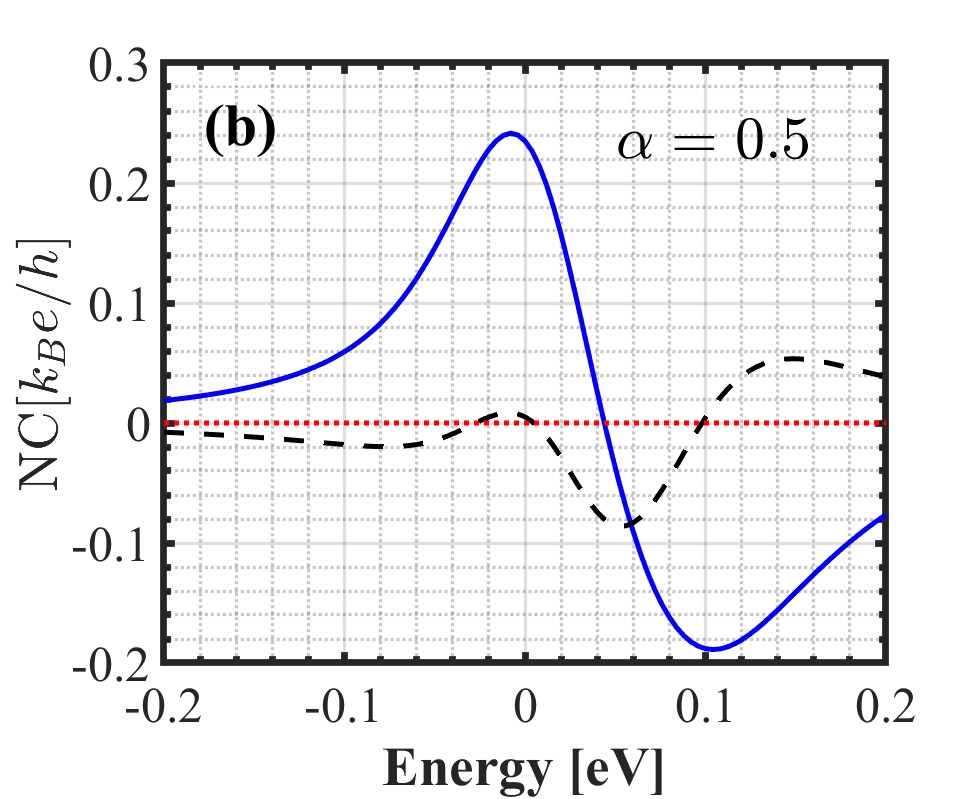}\\[2\tabcolsep]
		\includegraphics[width=0.5\linewidth]{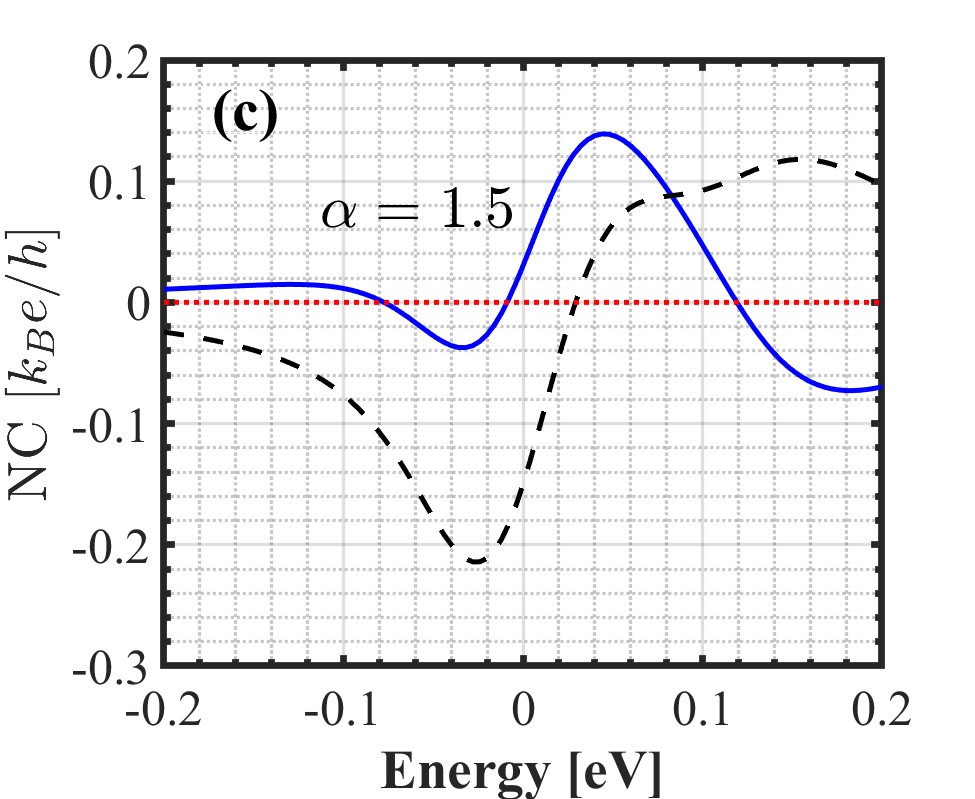}&
		\includegraphics[width=0.5\linewidth]{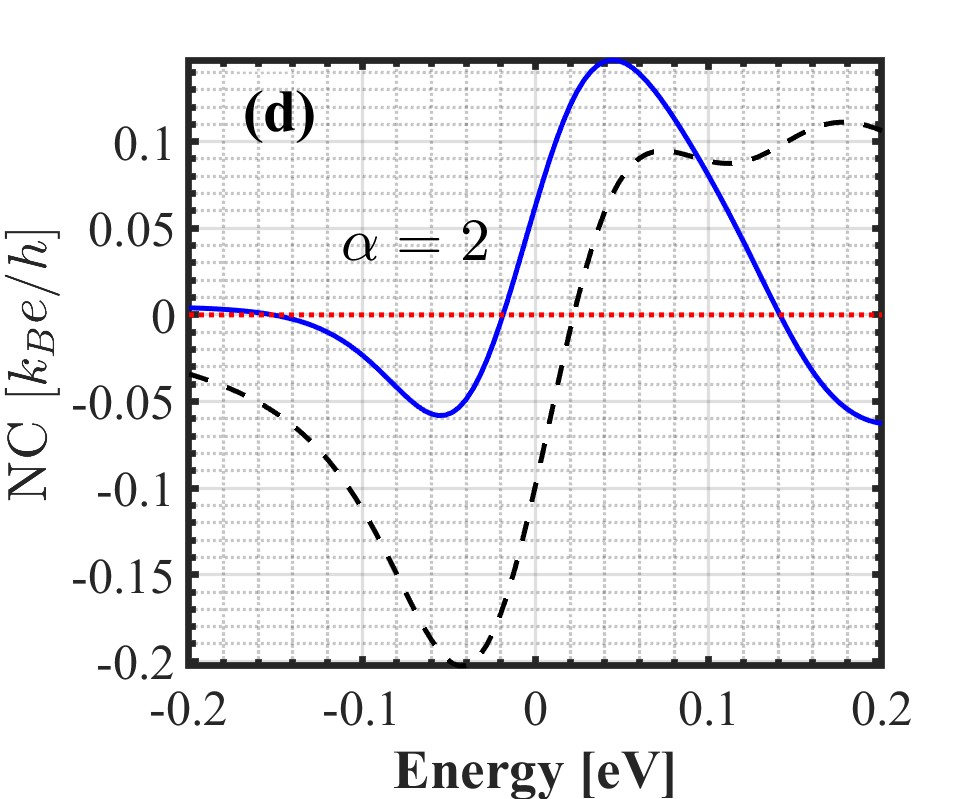}
	\end{tabular}
	\caption{The spin (SNC), valley (VNC), and total Nernst
		coefficient (TNC) versus the Fermi energy at $T = 300$ K for different
		values of $\alpha$. The blue, black, and red curves show SNC, VNC and
		TNC respectively.}\label{fig:7}
\end{figure}

\section{Results and Discussion\label{sec:3}} 

Figures~\ref{fig:2} (a) and (b) illustrate the spin texture in momentum space for the conduction and valence bands respectively. The color of the arrows shows the magnitude of the spin, $s=\sqrt{s_x^2+s_y^2}$ which varies from 0 (blue) to 1 (red). The tangential alignment of spins with momentum indicates spin-momentum locking, a signature of systems with topological properties. The absence of spin at the $\vec{k}=(0,0)$ suggests a Dirac-like point, where the energy bands cross, forming a spin-degenerate state that is the hallmark of the Dirac system. Elongating spin texture in the $k_y$ direction implies anisotropy in the band structure which is evident in Fig.\ref{fig:1} (c) and (d). Indeed, obviously there is a mirror symmetry with respect to the $\vec{k}=(0, 0)$ which is due to presence of the time-reversal symmetry in the system.  \par

The behavior of Berry curvatures around the Dirac points for various values of $\alpha$ and different bands is depicted in Figs.~\ref{fig:3} (a-d). Here, $\lambda=1, 2$  correspond to the valence and conduction bands respectively. In all scenarios, $\Omega_{\kappa,s}^n(\vec{k})$ is predominantly localized near the $K$ and $K^{\prime}$ valleys. Large peaks near the Dirac points indicate strong topological effects at these points, which could result in phenomena like anomalous velocity, orbital magnetization, or spin Hall effects. For $s=\uparrow$ and $\alpha<1$ (with $\kappa=\pm$), the sign of the conduction Berry curvature is positive (and negative for the valence band). However, for $\alpha>1$, the signs of the Berry curvatures are reversed. This alteration in the sign of the Berry curvature for different amounts of $\alpha$ indicates a topological phase transition in 1T$^{\prime}$-MoS$_2$. The opposite signs for the conduction and valence bands suggest a contribution to opposite topological currents from electrons in these bands, which could cancel or add under specific conditions. In contrast, for $s=\downarrow$, the signs of the Berry curvatures for both the conduction and valence bands are opposite to those observed for $s=\uparrow$. The opposing signs of the Berry curvature suggest that the material could host a nontrivial topology, potentially enabling edge states or chiral transport which is key in materials with nonzero Chern numbers or topological invariants.\par

The spin-valley-resolved Hall conductivity is illustrated in Fig.~\ref{fig:4} for various values of $\alpha$. When $\alpha<1$, the spin Hall conductivity reaches maximum values of $\pm1$ for spin-up and spin-down, respectively, while the valley Hall conductivity is zero (see Fig.~\ref{fig:4} (a) and (b)) and consequently $C_v=0$ and $C_s=1$ which shows QSHI phase. In this scenario, the energy range where the Hall conductivity remains constant indicates a spin-orbit coupling gap of about $0.04eV$. As shown in Fig.~\ref{fig:4} (b) and Fig.~\ref{fig:4} (c), when we increase $\alpha$ to higher amounts (still $\alpha<1$), the valley Hall conductivity becomes non-zero; however, the spin Hall conductivity continues to dominate. This increase in $\alpha$ reduces the spin-orbit coupling gap, which is consistent with the observations in Fig.~\ref{fig:1} (b). The Hall conductivity for $\alpha>1$ is different in comparison with $\alpha<1$. In this case, the valley Hall conductivity is no-zero and is the dominant term. The maximum amounts of the Hall conductivity are $\pm1$ which are related to the $\sigma_{\mp}$ respectively and $C_v=1$. This change in the Chern numbers for $\alpha>1$ is an evidence for the topological phase transition between QSHI-to-BI which is consistent with previous study~\cite{PhysRevB.103.125425}.\par

To cover the entire range of $V$ and Fermi energy, we plotted the color density of the spin-valley-resolved Hall conductivity as a function of $V$ and Fermi energy for both $\alpha<1$ and $\alpha>1$  (see Fig.~\ref{fig:5} ). For $\alpha<1$ ($\alpha >1$), the valley Hall conductivity (spin Hall conductivity) is zero, so we omitted these zero terms from our plots. A prominent diagonal bright region (white-yellow) extends linearly through the origin (see Fig.~\ref{fig:5} (a)). The slope of this bright region suggests a direct relationship between $V$ and $E_F$, meaning changes in the gate voltage correspond to shifts in the Fermi energy. Away from the central bright stripe, the conductivity rapidly decreases, as evidenced by the transition to darker shades (red to black). This indicates that the Hall response is sharply localized in the $V-E_F$ space. This region indicates a peak in the Hall conductivity ($\sigma_{\uparrow}=1$) that aligns with specific combinations of $V$ and $E_F$. As $E_F$ moves away from the band edges, the Hall conductivity decreases. This behavior is consistent with the vanishing Berry curvature in regions of
$k$-space far from band crossings, where the bands are less topologically active. As $V$ increases, this maximum shifts towards higher Fermi energies, and it is clear that the width of the light region remains constant, indicating that the topological gap is independent of the $V$ amount. A similar condition applies to $\sigma_{\downarrow}$  for $\alpha <1$ across different values of $V$, as shown in Fig.~\ref{fig:5} (b). The maximum value of  $\sigma_{\downarrow}$ is -1, which is located at the same position as that of $\sigma_{\uparrow}$. Since for $\alpha<1$, the valley Hall conductivity is zero, we only plotted it for $\alpha>1$ (see Figs.~\ref{fig:5} (c) and (d)). In this scenario, the maximum value of the Valley Hall conductivity is $\sigma_{+}= -1$ which is localized at certain combinations of $V$ and $E_F$. As $V$ increases (decreases), the Fermi energy interval shifts to higher (lower) energies. The valley Hall conductivity for $\sigma_{-}$ reaches a maximum value of +1 and exhibits behavior similar to that of $\sigma_{+}$ (Fig~\ref{fig:5} (d)).\par

We plotted the Chern number as a function of $V$ for varying values of $\alpha$ at $E_F=0$ and $T=0$. Figure.~\ref{fig:6} (a), shows the Chern numbers for $\alpha<1$. The valley Hall conductivity remains zero across all $V$ values. However, in the interval $-0.07 \leq V \leq 0.03$, we find a non-zero spin Hall conductivity with $C_{\uparrow}=1$, $C_{\downarrow}=-1$ and $C_s=1$, indicating the presence of a QSHI phase. In this case, both spins have the helical edge mode. For values of $V \leq -0.07$ and $V \geq 0.03$, the Hall conductivity vanishes, resulting in a trivial topological phase (TTP). However, when $\alpha > 1$, a topological phase transition occurs from QSHI-to-BI, characterized by $C_{\uparrow, \downarrow}=0$, $C_+=-1$, and $C_-=1$ (see Fig.~\ref{fig:6} (b)). In this case, $C_s=0$ and $C_v=1$, which leads to the absence of the spin edge modes. Notably, for $\alpha>1$ there is a wider range of $V$ values where the Hall conductivity is non-zero resulting in wider non-trivial region (see Fig.~\ref{fig:6} (c)). It is necessary to mention that when the Fermi energy falls within a band ($\alpha=1$), the Chern number is not defined, however, the Kubo formula can still provide the Hall conductivity.\par

We calculated the spin Nernst (SNC), valley Nernst (VNC), and total Nernst coefficients (TNC) as functions of the Fermi energy at room temperature for various values of $\alpha$. As shown in Fig.~\ref{fig:7} (a-d). For all values of $\alpha$, the TNC remains zero across the energy range, indicating a strong cancellation between the spin and valley contributions. When $\alpha=0$ (see Fig.~\ref{fig:7} (a)), the only non-zero term is SNC. The SNC exhibits a clear asymmetry around zero energy, with a significant negative (positive) peak at positive (negative) energies. This behavior suggests that the spin-related transport properties vary with the energy landscape. As an electric field is applied, the VNC begins to increase; however, SNC still dominates (Fig.~\ref{fig:7} (b)). When $\alpha$ is increased to 1.5 and 2 (see Figs.~\ref{fig:7} (c) and (d)), the VNC has a pronounced negative peak at slightly negative energies and transitions to a less negative or slightly positive regime at higher energies. This indicates a significant valley-dependent response that is similar in sign to the spin contribution in certain regions. The spin Nernst effect is dominant in the positive energy range, where it exhibits a positive peak. On the other hand, the valley Nernst effect is dominant in the negative energy range, contributing to a negative peak. The zero total Nernst coefficient suggests that, while the individual contributions are large, their combined transport properties result in minimal net transverse voltage under a thermal gradient.

\section{Conclusions\label{sec:4}}

In conclusion, this study uncover the topological phases and properties of  the 1T$^{\prime}$-MoS$_2$ theoretically. The analysis of spin textures reveals hallmark features such as spin-momentum locking, anisotropy in the band structure, and the presence of Dirac points, underscoring their potential for spintronic applications. The Berry curvature and Hall conductivity analyses demonstrate that the system undergoes a topological phase transition from a QSHI-to-BI phase as the parameter $\alpha$ is varied, with distinct spin-resolved and valley-resolved contributions to the Hall conductivity. Furthermore, the behavior of the spin and valley Nernst coefficients reveals nuanced energy-dependent transport phenomena, where significant contributions from spin and valley effects largely cancel each other, resulting in a zero total Nernst coefficient.                             
      
\section*{Data availability}
The data that support the findings of this study are available from the corresponding author upon reasonable request.
\section*{Conflict of interest}
The authors have no conflicts to disclose.
\bibliography{bib}
\end{document}